\documentclass[prb,twocolumn]{revtex4}

\usepackage{amsmath,amsfonts}
\usepackage{graphicx}
\usepackage{enumerate}

\newcommand{\e}			{\mathrm{e}}
\newcommand{\dt}		{\mathrm{d}t}
\newcommand{\dw}		{\mathrm{d}w}
\newcommand{\dlambda}	{\mathrm{d}\lambda}
\newcommand{\tr}		{\mathrm{tr}}
\newcommand{\Tr}		{\mathrm{Tr}}

\begin{document}

% -------------------------------------------------------------------

\title
{Response of a Fermi gas to time-dependent perturbations: 
{R}iemann-{H}ilbert approach at non-zero temperatures}

\author
{Bernd Braunecker}

\affiliation
{Department of Physics, Brown University, Providence, Rhode Island 02912, USA}

\date{\today}


% -------------------------------------------------------------------

\begin{abstract}
We provide an exact finite temperature extension to the recently 
developed Riemann-Hilbert approach for the calculation of response
functions in nonadiabatically perturbed (multi-channel) Fermi
gases. We give a precise definition of the finite temperature 
Riemann-Hilbert problem and show that it is equivalent to a 
zero temperature problem. Using this equivalence, we discuss the 
solution of the nonequilibrium Fermi-edge singularity problem at 
finite temperatures.
\end{abstract}

% -------------------------------------------------------------------

\maketitle

% -----------------------------------------------------------------------------
\section{Introduction}

The response of a Fermi gas to an external, time-dependent perturbation
is a well-known hard problem. 
It has repeatedly been studied in various contexts, mostly on 
systems exhibiting the Orthogonality Catastrophe, Fermi-edge 
singularities, or the Kondo effect\cite{oc_fes}.
The technical difficulties are connected to the necessity of summing
an infinite number of logarithmically divergent diagrams, and 
has led to the development of a multitude of methods to tackle this
task, including diagram summations, singular integral equation
approaches, bosonization, and summation schemes based on Slater
determinants\cite{oc_fes}. Many of these approaches are technically
involved or require assumptions such as, for instance, 
the separability of the external potential. 
In addition, the recent controversies 
concerning nonequilibrium Fermi systems\cite{neq_prim}
have shown that extensions of these methods are problematic and 
hard to achieve. A unified approach, depending on a minimum of
assumptions is, therefore, desirable.

Very recently, a major step in this direction has been made in the
form of the 
\emph{Riemann-Hilbert approach}\cite{Muzykantskii03a,dAmbrumenil05},
and applied to the calculation of Fermi-edge singularities at $T=0$ in 
Fermi gases under nonequilibrium constraints\cite{Muzykantskii03b}.
In this approach, the calculation of general response functions
in the perturbed Fermi gas is transformed into the problem 
of solving a
(generally non-abelian) Riemann-Hilbert (RH) boundary value
problem: The response functions are rewritten in terms of complex
matrix functions $Y(t)$ that satisfy boundary conditions, 
for $\mathrm{Im} t \to \pm 0$, expressed by a matrix function $R(t)$. 
In most cases of interest, $R(t)$ is directly related to the 
scattering matrix of the time-dependent potential.

In this paper, we extend the RH technique to finite temperatures.
We first give a precise definition of the finite temperature 
RH problem. 
We then show that the finite temperature problem is  
equivalent to a different zero temperature RH problem through the bijective 
mapping of the time variable $t \mapsto \tanh(\tfrac{\pi}{\beta}t)$,
with $\beta = 1/k_B T$. In simple 
(i.e. abelian) cases, this allows us to construct the explicit 
solution, but generally the resulting RH problem remains
highly nontrivial. We show, however, on the concrete example
of the nonequilibrium problem of
Refs.~\onlinecite{Muzykantskii03a,Muzykantskii03b,dAmbrumenil05}
how the equivalence between finite and zero temperature solutions
can be explored to construct approximate solutions to the
corresponding RH problem in the high and low temperature limits.

The paper is organized as follows: In Sec.~\ref{sec:zerotemp}, we
summarize the essential steps in the zero temperature RH approach.
In Sec.~\ref{sec:finitetemp}, the finite temperature RH problem is
defined, and the equivalence to the zero temperature case is shown.
Finally, in Sec.~\ref{sec:discussion}, we discuss the finite
temperature extensions for some simpler examples and for 
the nonequilibrium Fermi-edge singularity problem discussed in 
Refs.~\onlinecite{Muzykantskii03a,Muzykantskii03b,dAmbrumenil05}.

% -----------------------------------------------------------------------------
\section{Zero temperature case}
\label{sec:zerotemp}

We start with the introduction of the required concepts and notations
for the zero temperature RH approach, following Ref.~\onlinecite{dAmbrumenil05}.

The approach has been developed for the calculation of response functions
in a nonadiabatically perturbed system of fermions. We consider a 
gas of noninteracting fermions that is exposed to a time-dependent
potential $M$. The Hamiltonian is assumed to take the form
\begin{equation}
\begin{aligned}
	\hat{H}(t) &= \hat{H}_0 + \sum_{\epsilon,\epsilon'} 
	\mathbf{a}_\epsilon^\dagger M(t,\epsilon,\epsilon') \mathbf{a}_{\epsilon'},
	\\
	\hat{H}_0  
	&= \sum_\epsilon \epsilon \, 
	\mathbf{a}_\epsilon^\dagger \mathbf{a}_\epsilon,
\end{aligned}
\end{equation}
where $\epsilon$ labels the single-particle energies of the electrons, and
$\mathbf{a}_\epsilon^\dagger$ is a vector of the electron operators 
$a_{\epsilon,j}^\dagger$, where $j=1,\dots,N$ is a channel index labeling 
any further classification of the single particle states.
The potential $M(t,\epsilon,\epsilon)$ is a $N\times N$ matrix in channel 
space, which we assume to be nonzero only for some finite time
interval $0<t<t_f$, during which the external perturbation is switched on.

The quantities of interest are response functions of the type
\begin{equation}
	\chi_R = \langle \hat{R} \rangle.
\end{equation}
At zero temperature, $\langle ... \rangle$ denotes the average over the 
ground state $|0\rangle$; at finite temperatures, as considered in 
Sec.~\ref{sec:finitetemp}, it represents 
the thermal average over the noninteracting system states at $t=0$.
The operator $\hat{R}$ is related to the time-evolution operators 
$\hat{U}(t_f)$ and $\hat{U}_0(t_f)$ of $\hat{H}(t)$ and $\hat{H}_0$, 
respectively, or equivalently to the scattering matrix $S$ for the 
potential $M$.
For the present discussion, the precise form of $\hat{R}$ is unimportant.
A summary of necessary conditions on $\hat{R}$ is given in 
Appendix~\ref{sec:time_repr_R}, while a detailed discussion can be found 
in Refs.~\onlinecite{Muzykantskii03a,dAmbrumenil05} and \onlinecite{Adamov01};
some examples are given at the end of this paper. 

Since the Hamiltonian is quadratic in the electron operators, the 
action of the many-body operator $\hat{R}$ is fully specified by 
its action on single-particle states (see the appendices). 
Let $R = \{R_{j,j'}(\epsilon,\epsilon')\}$ be the matrix formed of the
matrix elements of $\hat{R}$ between the single-particle states $(j,\epsilon)$,
$R_{j,j'}(\epsilon,\epsilon') = \langle| a_{\epsilon,j} \hat{R} a_{\epsilon',j'}|\rangle$,
with $|\rangle$ the true vacuum.
For the following treatment, it is essential to assume that $R$ is diagonal
in the time representation, $R = R(t)$ (with $t$ the variable conjugated 
to the energy representation $\epsilon$).
For the response functions relevant for Fermi-edge singularity problems, 
it can be shown\cite{Muzykantskii03a,dAmbrumenil05,Adamov01} (see Appendix~\ref{sec:time_repr_R})
that this assumption is valid as long as the variation of the scattering matrix is slow 
compared to the delay time of the scattering process itself.

For the noninteracting system, $|0\rangle$ is a Slater determinant 
in the single particle states. As shown in Appendix~\ref{sec:deriv_eq3}, 
we can write
\begin{equation} \label{eq:det_chi}
	\chi_R = \det\bigl( (1-f) + f R \bigr),
\end{equation}
where $R$ represents the matrix elements of $\hat{R}$ taken between 
the single-particle states, and $f$ is the Fermi function, which has 
the matrix elements
\begin{equation}
	f_{j j'}(\epsilon, \epsilon') = \delta_{\epsilon  \epsilon'} \delta_{j j'}
	\theta(-\epsilon).
\end{equation}
In this expression, we have set the chemical potential 
$\mu$ to zero. This is possible
without loss of generality even in the nonequilibrium case of several 
Fermi seas $j$ with different chemical potentials $\mu_j$,
such as for biased tunneling barriers. For the unperturbed 
system with uncoupled Fermi seas, we can shift the single-particle
energies by $\epsilon \to \epsilon - \mu_j$ and set all 
chemical potentials to zero. 
In other words, the fermion operators are replaced by
$a_{\epsilon,j} \to \e^{i \mu_j t} a_{\epsilon,j}$.
The Hamiltonian keeps the same form except that the potential
$M(t)$, and thus $R(t)$, acquires an additional dependence 
on time\cite{Muzykantskii03a},
$R_{j,j'}(t) \to R_{j,j'}(t) \e^{i(\mu_j-\mu_{j'})t}$.

The logarithm of the determinant in Eq.~\eqref{eq:det_chi}
can be written in the form
\begin{equation} 
\begin{aligned}
	\ln\chi_R 
	&= \Tr[f \ln R] + \Tr[\ln(1-f+fR) - f \ln R]
	\\
	&\equiv \ln\chi_R^{(1)} + \ln\chi_R^{(2)},
\end{aligned}
\end{equation}
with $\Tr$ being the trace over channel indices $j$ and energy.
$\ln\chi_R^{(1)} = \Tr[f \ln R]$ consists of the diagonal terms 
of $\ln R$ summed over the occupied states of the unperturbed system.
Depending on the system under investigation, it expresses, for 
instance, a shift of the ground state or threshold energy 
(Fumi's theorem\cite{Fumi55} if $R$ is the scattering matrix) together 
with an imaginary correction for nonequilibrium 
conditions\cite{Muzykantskii03b}, or the average transfer of charge 
across a barrier\cite{Muzykantskii03a}.
The second term, $\ln\chi_R^{(2)}$, contains the nontrivial effects 
close to the Fermi surface which are, for instance, associated with 
Fermi-edge singularities. In the following, we will focus on 
$\ln\chi_R^{(2)}$ only.

It is convenient to switch from the energy representation to the time
representation of the trace, in which $R$ is diagonal.
The Fermi function, however, becomes nondiagonal in time and reads
\begin{equation}
	f_{j j'}(t,t') = \frac{i/2\pi}{t-t'+i0} \delta_{j j'}.
\end{equation}
(Here and henceforth we set $\hbar = 1$.)
The logarithm of an infinite matrix is represented by introducing a
dependence on a parameter $\lambda$ as
\begin{equation}
	R \to R(\lambda) = \exp(\lambda \ln R)
\end{equation}
and by writing
\begin{equation} \label{eq:chi_R0}
	\ln\chi_R^{(2)} = \int_0^1 \mathrm{d}\lambda \int \dt
	\tr\left[ \left[ (1-f+fR)^{-1} f - f R^{-1}\right] 
	\frac{\mathrm{d}R}{\mathrm{d}\lambda}
	\right],
\end{equation}
where $\tr$ denotes the trace over the channel index $j$ only.

The difficult task is to find the inverse of $(1-f+fR)$.
For this, a complex $N\times N$ matrix function $Y(t)$ is introduced,
which solves an auxiliary \emph{RH problem}, defined by:
\begin{enumerate}[(a)]
\item \label{prop:1}
$Y(t)$ is a nonsingular $N \times N$ matrix function, which is analytic 
and nonzero in the complex plane, except on the interval $(0,t_f)$,
where it satisfies the boundary condition
\begin{equation} \label{eq:RH}
	Y_-(t) Y_+^{-1}(t) = R(t),
\end{equation}
for $Y_\pm(t) = Y(t\pm i 0)$, and except at the end points $t_e=0,t_f$,
where $Y(t)$ may vanish or be weakly singular as 
$|Y(t)| \sim  |t-t_e|^\varepsilon$ for $-1 < \varepsilon < 1, \varepsilon \neq 0$.

\item \label{prop:2}
$Y(t)$ has the asymptotic condition
\begin{equation} \label{eq:asympY}
 	Y(t) \to 1 \qquad \text{when $|t| \to \infty$}.
\end{equation}
\end{enumerate}
From these two properties, two important identities follow:
\begin{align}
	f Y_+ f &= Y_+ f, \label{eq:id+}\\
	f Y_- f &= f Y_-, \label{eq:id-}
\end{align}
which equally hold if $Y$ is replaced by $Y^{-1}$.
In the time representation, Eq.~\eqref{eq:id+} reads
\begin{equation} \label{eq:int_zero_temp}
	\int \dt \frac{i/2\pi}{t_1-t+i0} Y_+(t) \frac{i/2\pi}{t-t_2+i0}
	=  Y_+(t_1) \frac{i/2\pi}{t_1-t_2+i0},
\end{equation}
which is proved by closing the contour in the upper half-plane.
A similar contour in the lower half-plane proves 
Eq.~\eqref{eq:id-}.

In terms of the functions $Y_\pm$, the inverse of $(1-f+fR)$ is expressed by
\begin{equation} \label{eq:inverse}
	\bigl(1-f+fR\bigr)^{-1} = Y_+ \bigl[ (1-f)Y_+^{-1} + f Y_-^{-1} \bigr],
\end{equation}
which can be verified by multiplication by $(1-f+fR)$ from the right
or left, and by making use of the identities \eqref{eq:id+} and \eqref{eq:id-}.

Finally, the following result 
\begin{align}
	\int \dt \, \tr\bigl\{[A,f] B\bigr\} 
	&= \int \dt \lim_{t'\to t} \tr \frac{i}{2\pi} 
	\left[\frac{A(t)-A(t')}{t-t'+i0}\right] B(t')
	\nonumber \\
	&= \frac{i}{2\pi} \int \dt \,
		\tr \left\{\frac{\mathrm{d}A(t)}{\dt} B(t) \right\}
	\label{eq:relation}
\end{align}
for any differentiable matrix functions $A$ and $B$,
allows to write Eq.~\eqref{eq:chi_R0} in the compact form
\begin{equation} \label{eq:sol}
	\ln\chi_R^{(2)} = 
	\frac{i}{2\pi} \int_0^1 \mathrm{d}\lambda \int \dt \,
	\tr\left\{ \frac{\mathrm{d}Y_+}{\dt} Y_+^{-1}
	R^{-1} \frac{\mathrm{d}R}{\mathrm{d}\lambda} 
	\right\}.
\end{equation}
The response function $\chi_R^{(2)}$, therefore, is fully expressed in terms 
of the solution $Y(t)$ of the RH problem.

% -----------------------------------------------------------------------------
\section{Finite temperature case}
\label{sec:finitetemp}

We show in this section, that we can define a finite temperature version
of the RH problem such that the important formula \eqref{eq:inverse}
(and, therefore, Eq.~\eqref{eq:sol}) remains valid. 
Subsequently, we show that this new RH problem
is equivalent to a zero temperature RH problem.

\subsection{Finite temperature RH problem}

Note that Eq.~\eqref{eq:inverse} remains valid if we generalize the
identities \eqref{eq:id+} and \eqref{eq:id-} to 
\begin{align}
	f Y_+ f &= Y_+ f + F, \label{eq:id+1}\\
	f Y_- f &= f Y_- + F, \label{eq:id-1}
\end{align}
with the same $F = F(t,t')$, a (finite) matrix function, appearing in both 
identities.

At finite temperatures $T$, the Fermi function becomes 
\begin{equation}
	f(t,t') = \frac{i/2\pi}{\frac{\beta}{\pi} 
	\sinh(\frac{\pi}{\beta}(t-t')+i0)},
\end{equation}
with $\beta = 1/k_B T$.
With this function, Eq.~\eqref{eq:id+1} reads explicitly
\begin{multline} \label{eq:id+2}
	\int_{-\infty}^\infty \dt 
	\frac{i/2\beta}{\sinh\bigl(\frac{\pi}{\beta}(t_1-t+i0)\bigr)}
	Y_+(t)
	\frac{i/2\beta}{\sinh\bigl(\frac{\pi}{\beta}(t-t_2+i0)\bigr)}
	\\ 
	= Y_+(t_1) 
	\frac{i/2\beta}{\sinh\bigl(\frac{\pi}{\beta}(t_1-t_2+i0)\bigr)}
	+ F(t_1,t_2).
\end{multline}
The Fermi function is antiperiodic in the imaginary time direction,
$f(t+i\beta,t') =  -f(t,t')$. If we impose that Eq.~\eqref{eq:id+1} 
and \eqref{eq:id-1} are invariant under such transformations,
$Y(t)$ and $F(t,t')$ must be (anti)periodic in the imaginary 
time direction as $Y(t+i\beta) = Y(t)$ and 
$F(t+i\beta,t')=F(t,t'+i\beta) =-F(t,t')$.
This allows us to restrict the analysis to the
strip $-\beta/2 < \mathrm{Im}\, t < \beta/2$. 

The \emph{finite temperature RH problem} can then be formulated as follows:
\begin{enumerate}[(i)]
\item \label{prop:i}
$Y(t)$ is a nonsingular $N \times N$ matrix function, which 
is analytic in the strip 
$\mathcal{S}_\beta = \{ t \in \mathbb{C} \, | \, 
 -\beta/2 < \mathrm{Im}\, t < \beta/2 \}$,
except on the interval $(0,t_f)$, where it satisfies
Eq.~\eqref{eq:RH}, and except at the end points $t_e =0,t_f$,
where it may vanish or be weakly singular as 
$|Y(t)| \sim |t-t_e|^\varepsilon$ with $-1< \varepsilon < 1, \varepsilon \neq 0$.
\item \label{prop:ii}
$Y(t)$ tends to definite finite values as $|t| \to \infty$ in 
$\mathcal{S}_\beta$
(the values on the far right or far left, however, are generally different).
\item \label{prop:iii}
$Y(t)$ can be analytically continued through 
$t \pm i \beta/\pi$, for real $t$, such that
$Y(t+i\beta/2) = Y(t-i\beta/2)$.
For convenience, we normalize $Y(t)$ (by multiplication by a constant
matrix from the right) such that $Y(\pm i \beta/2) = 1$.
\end{enumerate}
Condition \eqref{prop:iii} is absent in the zero temperature
case, but is essential for the validity of
Eqs.~\eqref{eq:id+1} and \eqref{eq:id-1}.

\begin{figure}
\begin{center}
	\includegraphics[width=\columnwidth]{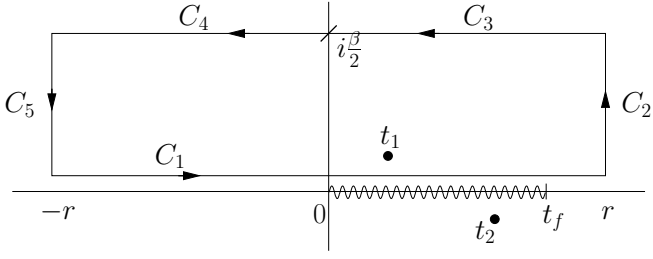}
	\caption{Integration contour in the $t$-plane. 
	The wiggled line indicates the branch cut for $Y(t)$,
	the isolated dots the poles of the integrand.
	The limit $r \to \infty$ is assumed.
	\label{fig:contour1}}
\end{center}
\end{figure}

Indeed, let us first focus on Eq.~\eqref{eq:id+1}. We integrate $f Y_+ f$ 
over the contour $C = C_1+ \dots +C_5$ shown in Fig.~\ref{fig:contour1}. 
In the figure, $r$ is a large number, which we eventually send to infinity.
Since $Y(t)$ remains bounded as $|t|\to\infty$, the integrals over the vertical lines 
$C_2$ and $C_5$ vanish in this limit, and we obtain Eq.~\eqref{eq:id+1} 
with $F$ given by the integral of $f Y_+ f$ 
over the line through $i\beta/2$, parallel to the real axis. 

A similar contour in the lower part of $\mathcal{S}_\beta$ leads to 
Eq.~\eqref{eq:id-1}, with $F$ given by the integral over the line
through $-i\beta/2$, parallel to the real axis. Both integrals for $F$
are identical due to the periodicity and analyticity of $Y(t)$, 
and Eqs.~\eqref{eq:id+1} and \eqref{eq:id-1} are fulfilled.

The solution of the finite temperature RH problem, together 
with Eqs.~\eqref{eq:id+1} and \eqref{eq:id-1}, immediately
leads to the solution for the response function $\chi_R$ by means
of Eq.~\eqref{eq:sol}. (Note that Eq.~\eqref{eq:relation} remains
true when replacing $(t-t')^{-1}$ by $[(\beta/\pi)\sinh(\pi (t-t') /\beta)]^{-1}$.)

\subsection{Equivalence to zero temperature case}

The finite temperature RH problem, defined by conditions 
\eqref{prop:i}-\eqref{prop:iii}, is equivalent to the RH
problem at zero temperature, defined by conditions
\eqref{prop:1} and \eqref{prop:2}, upon a redefinition
of the boundary condition \eqref{eq:RH}.

The equivalence is shown by means of the mapping
\begin{equation} \label{eq:mapping}
	t \mapsto w = \tanh(\tfrac{\pi}{\beta}t),
\end{equation}
which is often used to switch between zero and finite temperatures,
and which describes a bijection between the strip $\mathcal{S}_\beta$
and the slit plane 
$\mathcal{S} = \{ w \in \mathbb{C} | w \not\in (-\infty,-1], w \not\in [1,\infty) \}$.
Let us set $\tilde{Y}(w) = Y(t)$ for $w = \tanh(\tfrac{\pi}{\beta}t)$.
\begin{itemize}
\item
The boundary condition \eqref{eq:RH} for $Y(t)$ and $t \in (0,t_f)$
becomes a boundary condition of the same type \eqref{eq:RH} for 
$\tilde{Y}(w)$ and $w \in (0,w_f)$, 
$0 < w_f = \tanh(\tfrac{\pi}{\beta}t_f) < 1$, and the matrix
$\tilde{R}(w) = R(t)$.
\item
The analyticity of $Y(t)$ in $\mathcal{S}_\beta\backslash[0,t_f]$ ensures the
analyticity of $\tilde{Y}(w)$ in $\mathcal{S}\backslash[0,w_f]$. 
The periodicity $Y(t+i\beta/2) = Y(t-i\beta/2)$ for real $t$ is nothing 
but the condition of analyticity of $\tilde{Y}(w)$ through the real axis 
at $|w|>1$.
\item
$t = \pm i\beta/2$ maps onto infinity in the $w-$plane. With the 
normalization $Y(\pm i \beta/2)=1$, we have $\tilde{Y}(w) \to 1$
as $|w| \to \infty$. On the other hand, the boundedness of $Y(t)$
as $|t| \to \infty$ in $\mathcal{S}_\beta$ implies that
$\tilde{Y}(w)$ is finite at $w = \pm 1$.
\end{itemize}
This proves that the zero and finite temperature versions of the RH
problem are equivalent, upon the change of boundary condition
$R(t) \to \tilde{R}(w) = R(t)$.

With the mapping, the identities \eqref{eq:id+1} and \eqref{eq:id-1} 
are mapped onto \eqref{eq:id+} and \eqref{eq:id-}, too. 
The infinite $t$-integrals over the real axis, map onto $(-1,1)$ 
in the $w$-plane (including the cut $[0,w_f]$), while the integrals
over the contours $C_3$ and $C_4$ provide the remaining parts of 
the real axis in the $w$-plane, $(1,\infty)$ and $(-\infty,-1)$, respectively,
at which $\tilde{Y}(w)$ is analytic. The integrands can be identified
through the formula
\begin{equation} \label{eq:sinh_formula}
	\sinh\bigl(\tfrac{\pi}{\beta}(t-t')\bigr) 
	= \cosh(\tfrac{\pi}{\beta}t) \cosh(\tfrac{\pi}{\beta}t')
	  (w - w').
\end{equation}
The dependence on $\cosh(\tfrac{\pi}{\beta}t)$ in the integrand
cancels with $\frac{\dw}{\dt} = (\pi/\beta)/\cosh^2(\tfrac{\pi}{\beta}t)$.
After factorization of 
$(\pi/\beta)\cosh(\tfrac{\pi}{\beta}t_1)\cosh(\tfrac{\pi}{\beta}t_2)$, we 
see that Eqs.~\eqref{eq:id+1} and \eqref{eq:id-1} are identical to
\eqref{eq:id+} and \eqref{eq:id-}.

% -----------------------------------------------------------------------------
\section{Examples}
\label{sec:discussion}

A solution to the finite temperature RH problem allows the calculation of
the response function \eqref{eq:sol}. We illustrate this with a few simpler examples
of how the equivalence to the zero temperature RH problem can be used to
construct explicit finite temperature solutions. We then discuss approximate
finite temperature solutions for the nonequilibrium RH problem of 
Ref.~\onlinecite{Muzykantskii03b}.

\subsection{Some simpler examples}

The RH problem appeared early in the study of Fermi-edge
singularities. It is part of the solution of the Dyson equation
for the Green's function, formulated by Nozi\`eres and De~Dominicis 
as a singular integral equation\cite{ND69}, for which the standard method
of solution consists in the transformation into a RH problem\cite{singinteq}. 
In Nozi\`eres and De~Dominicis' case, the boundary condition $R$ is a
constant, for $t \in (0,t_f)$, given by the single-particle single-channel 
scattering matrix $R = S = \e^{-2i\delta}$, where $\delta$ is the scattering 
phase shift ($|\delta|<\pi/2$). 
The solution is given by
\begin{equation} \label{eq:RH_ND}
	Y(t) = \bigl((t-t_f)/t\bigr)^{\delta/\pi}.
\end{equation}
The finite temperature extension to this expression was achieved 
by Yuval and Anderson\cite{YA70}
in their application of the Nozi\`eres-De~Dominicis approach to the Kondo
problem. It resulted from writing $1/\sinh(\pi t/\beta)$, the free Green's 
function in the kernel of the integral equation, as the sum of 
$(-1)^n/(t+i n \beta)$ over integer $n$ in the kernel of the integral 
equation. The corresponding RH problem becomes the product of zero-temperature 
RH problems with branch cuts shifted by $i n \beta$. Hence, the finite 
temperature solution found by Yuval and Anderson reads
\begin{equation} \label{eq:RH_YA}
	Y(t) = \prod_n \left(\frac{t-t_f+i n \beta}{t+i n \beta}\right)^{\delta/\pi}
	= \left( \frac{\sinh(\frac{\pi}{\beta}(t-t_f))}{\sinh(\frac{\pi}{\beta}t)} 
	  \right)^{\delta/\pi}.
\end{equation}
With the replacement $t \mapsto \tanh(\tfrac{\pi}{\beta}t)$ in 
Eq.~\eqref{eq:RH_ND} and the use of Eq.~\eqref{eq:sinh_formula}, 
this solution can immediately be verified, as constant matrices 
$R$ and $\tilde{R}$ coincide.
Note that with the mapping we obtain an additional constant factor 
$\cosh(\tfrac{\pi}{\beta}t_f)^{\delta/\pi}$, reflecting the fact that 
a solution of the RH problem remains a solution upon multiplication by a
constant matrix from the right. With our normalization \eqref{prop:iii}
this factor is suppressed. Such constants are unimportant since 
they drop out in the physical solution \eqref{eq:sol}. 

Eqs.~\eqref{eq:RH_ND} and \eqref{eq:RH_YA} remain valid if $R$ is a time-independent
diagonalizable matrix. In this case, $\delta$ becomes a matrix with eigenvalues representing
the phase shifts of the multi-channel scattering matrix $R$.

For scalar, but time-dependent $R(t)$, the solution of the RH problem 
can be constructed in a standard way\cite{singinteq}, and is 
given by
\begin{equation} 
	Y(t) = \exp\left(\frac{1}{2\pi i} \int_0^{t_f} \dt'
	                 \frac{\ln R(t')}{t'-t} \right).
\end{equation}
At finite temperatures, the mapping allows here the explicit solution, too,
which has the same form as the previous equation, with $R$ replaced by 
$\tilde{R}$,
\begin{equation}
\begin{aligned}
	Y(t) 
	&= \exp\left(
		\frac{1}{2\pi i} \int_0^{\tanh(\frac{\pi}{\beta}t_f)} \dw'
		\frac{\ln \tilde{R}(w')}{w'-\tanh(\frac{\pi}{\beta}t)}
	   \right) 
	\\
	&= \exp\left(
		\frac{1}{2\beta i} \int_0^{t_f} \dt' 
		\frac{\cosh(\frac{\pi}{\beta} t)}{\cosh(\frac{\pi}{\beta}t')}
	    \frac{\ln R(t')}{\sinh(\frac{\pi}{\beta}(t'-t))} \right).
\end{aligned}
\end{equation}
Notice that all these expressions become singular at $t\to 0$ or
$t \to t_f$. These divergences are spurious and due to the approximation
of an infinite bandwidth (given by the form of the Fermi function), and
have a natural cutoff in the true inverse bandwidth\cite{ND69}, 
$1/\xi$.

In these simple examples, the response function \eqref{eq:sol} 
can explicitly be calculated. 
Let us, for instance, consider Eq.~\eqref{eq:RH_YA}.
Since $S^{-1} \frac{\mathrm{d} S}{\dlambda} = -2 i \delta$, we obtain
\begin{multline} \label{eq:sol_eq}
	\ln \chi_R^{(2)}(t_f) 
	= 
	\frac{\tr\{\delta^2\}}{2\pi^2}
	\int_0^{t_f} \dt 
	\frac{\mathrm{d}}{\dt} 
	\ln\left[
		\frac{\sinh(\frac{\pi}{\beta}(t-t_f+i0))}{\sinh(\frac{\pi}{\beta}t+i0)} 
	\right]
	\\
	=
	-
	\frac{\tr\{\delta^2\}}{\pi^2}
	\ln\left[\frac{\sinh(\pi t_f/\beta)}{\sinh(\pi/\beta\xi)}\right],
\end{multline}
again with the cutoff at the inverse bandwidth $1/\xi$.
For short times $t_f \ll \beta$, we obtain a power-law similar to the
well known zero temperature result 
$\chi_R^{(2)}(t_f) \sim t_f^{-\delta^2/\pi^2}$,
which, however, crosses over into an exponential decay as $t_f \gg \beta$,
$\chi_R^{(2)}(t_f) \sim \e^{- (\delta^2/\pi\beta) t_f}$.

\subsection{Biased tunneling barriers}

As a more involved example, 
let us consider the non-abelian RH problem, in which $R(t)$ is
a time-dependent matrix that does not allow a time-independent 
diagonalization.
Such situations naturally arise in nonequilibrium systems, and have recently 
been studied with the RH approach\cite{Muzykantskii03b,dAmbrumenil05}:
Consider, for instance, the coupling between a defect state in a tunneling
barrier and the connecting electrodes that are kept at a voltage
bias $V$. The defect is assumed to be either in its ground state $g$ 
or in in an excited state $e$. We further assume that in the ground state the 
defect is fully screened and does not interact with the conduction electrons,
whereas in the excited state it acts on the conduction electrons 
through the potential $M$. 
The diagonal matrix elements of this
potential describe the scattering of the electrons in 
the electrodes, while the nondiagonal matrix elements represent
a modification of the tunneling rate. The matrix $R=S$ is 
the scattering matrix for $M$. 
The response function $\chi_R(t=t_f-0)$ is related to the transition 
rate of the defect into the excited state, $p_{g\to e}(\Delta E)$,
by 
\begin{equation} \label{eq:trans_rate}
	p_{g\to e}(\Delta E) 
	= |u|^2 \int_0^\infty \dt \ \chi_R(t) \ \e^{i \Delta E t},
\end{equation}
which is the Golden Rule expression rewritten as a time integral.
$\Delta E$ is the bare energy difference between ground state
and excited state of the defect, and $u$ is the bare amplitude
in the Hamiltonian for the transition $g \to e$.

A closed solution to the resulting time-dependent RH problem is not known.
In the short time limit $t \ll 1/V$ and in the long time limit
$t \gg 1/V$, however, it is possible to construct approximate
solutions\cite{Muzykantskii03a,Muzykantskii03b,dAmbrumenil05}.
We first give an example for such an approximation at zero temperature,
and then discuss how it extends to finite temperatures.

\subsubsection{Zero temperature case}

In the nonequilibrium Fermi-edge singularity 
problem\cite{Muzykantskii03b,dAmbrumenil05}, 
$R(t)$ is a unitary time-dependent scattering matrix 
coupling two Fermi liquids, $j=1$ and $j=2$, at the left and at the 
right of a tunneling barrier. The nonequilibrium constraint is introduced
by a finite voltage bias $V$, keeping the left and right chemical 
potentials at the values $\mu_1=V+\mu$ and $\mu_2=\mu$, respectively
(the electronic charge $e$ is set to unity).
As noted in Sec.~\ref{sec:zerotemp}, we can shift $\mu_{1,2}$ to 0 in 
the unperturbed system by passing to the interaction representation
$\hat{H}_0 \to \tilde{H}_0 = \hat{H}_0 - \mu_1 \hat{N}_1 - \mu_2 \hat{N}_2$, 
where $\hat{N}_{1,2}$ are the number operators for the electrons in the 
electrodes 1 and 2. 
This is equivalent to a gauge transformation, transforming the electron 
operators as
$a_{\epsilon,j} \to \tilde{a}_{\epsilon,j} = \e^{i\mu_j t} a_{\epsilon,j}$.
The voltage difference now appears only as an additional time dependent phase
factor in the matrix elements of $R(t)$ that couple electrodes 1 and 2: 
\begin{equation} \label{eq:R(t)}
	R(t) = 
	\begin{pmatrix}
		R_{11}				& \e^{iVt} R_{12} \\
		\e^{-iVt} R_{21}	& R_{22}
	\end{pmatrix},
\end{equation}
where $R_{11}$ and $R_{22}$ are the backscattering matrix
elements, and $R_{12}$ and $R_{21}$ describe transmission 
through the barrier.
For $V=0$, the scattering matrix $R$ is time-independent during $(0,t_f)$,
and the explicit solution of the RH problem is given by
Eq.~\eqref{eq:RH_ND}, with $\delta$ the matrix of phase 
shifts, $R = \exp(-2 i\delta)$.

\begin{figure}
\begin{center}
\includegraphics[width=\columnwidth]{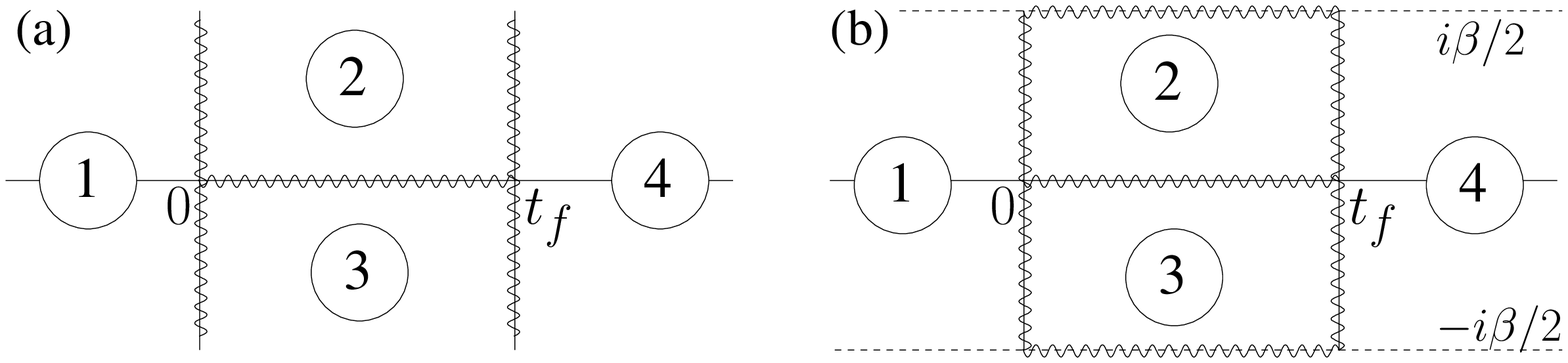}
\caption{Regions $s=1,\dots,4$ of definition of
the functions $W_j$ (Eq.~\eqref{eq:def_W}). Branch cuts are
indicated by the wiggled lines.
(a) Zero temperature case. (b) Modification for finite
temperatures with additional branch cuts at 
$\mathrm{Im} t = \pm \beta/2$.
\label{fig:zones}}
\end{center}
\end{figure}

If $V \neq 0$, an explicit solution to the RH problem is not known.
The naive solution, obtained by multiplication of Eq.~\eqref{eq:RH_ND}
by $V$-dependent phase factors $\exp(\pm i V t)$ would solve Eq.~\eqref{eq:RH} but 
violate condition \eqref{prop:2}, expressed by Eq.~\eqref{eq:asympY},
because of the divergence of $\exp(\pm i V t)$ for large complex-valued
arguments $t$. Such a divergence cannot be compensated by multiplication 
by $t$-dependent analytic matrix functions from the right. The 
solution \eqref{eq:sol} would become invalid.
An approximate solution, however, can be constructed as 
follows\cite{Muzykantskii03a,Muzykantskii03b,dAmbrumenil05}:
For $V>0$, we decompose $R(t)$ into the product
\begin{equation} \label{eq:decomp_R}
	R(t) = 
	\begin{pmatrix} 1 & 0 \\ \alpha/z & 1 \end{pmatrix}
	\begin{pmatrix} a & 0 \\ 0 & b \end{pmatrix}
	\begin{pmatrix} 1 & \alpha' z \\ 0 & 1 \end{pmatrix},
\end{equation}
with $z = \exp(i V t)$, $a = R_{11}$, $b=\det R/R_{11} = 1/R_{22}^*$, 
$\alpha = R_{21}/R_{11}$, and $\alpha' = R_{12}/R_{11}$. 
Now define the function
\begin{equation}
	\psi(t) = 
	\exp\left(\frac{\ln D}{2\pi i} 
	          \ln\left[\frac{t}{t-t_f}\right]\right),
\end{equation}
where $D = \mathrm{diag}(a,b)$. This function satisfies
the boundary condition
\begin{equation} \label{eq:psi_bc}
	\psi_-(t) \psi_+^{-1}(t) = D
\end{equation}
at $t \in (0,t_f)$.
To approximate the unknown solution $Y(t)$, we consider the following 
function $W(t)$:
\begin{equation} \label{eq:def_W}
\begin{aligned}
	W_1(t) &= \psi(t),
	&
	W_2(t) &= \begin{pmatrix} 1 & -\alpha' z \\ 0 & 1\end{pmatrix} \psi(t),
	\\
	W_3(t) &= \begin{pmatrix} 1 & 0 \\ \alpha/z & 1\end{pmatrix} \psi(t),
	&
	W_4(t) &= \psi(t), 
\end{aligned}
\end{equation}
where the functions $W_s$ are defined in the regions $s=1,\dots,4$ indicated in
Fig.~\ref{fig:zones} \mbox{(a)}. 
$W$ has branch cuts between the boundaries of all 4 regions $s$. Most 
importantly, between regions $s=2$ and $s=3$, it satisfies
\begin{equation} 
	W_3(t) W_2^{-1}(t) 
	=
	W_-(t) W_+^{-1}(t) 
	= R(t).
\end{equation}
Furthermore, $W(t) \to 1$ as $|t| \to \infty$. Hence
$W(t)$ satisfies the conditions \eqref{prop:1} and \eqref{prop:2}
of the RH problem, at the price of the additional vertical branch cuts
through $0$ and $t_f$. These branch cuts, however, become exponentially
small, $\propto \e^{- V |t|}$, with increasing distance to the real time
axis. For instance, consider the vertical cut at $\mathrm{Re}\,t=0$
in the upper half-plane. We have
\begin{equation} 
\begin{aligned}
	W_2(t) W_1^{-1}(t) 
	&= 
	\begin{pmatrix} 1 & -\alpha' \e^{-V |t|} \\ 0  & 1 \end{pmatrix}
	\psi(t)
	\psi^{-1}(t) 
	\\
	&=
	1 + O(\e^{-V |t|}).
\end{aligned}
\end{equation}
Similar exponential suppressions hold for the remaining vertical 
cuts. Therefore, $Y(t) \equiv W(t)$ is a good approximation
to the solution of the RH problem if we consider times $t$
such that $|t|, |t-t_f| \gg 1/V$.
Notice that in the boundary condition \eqref{eq:psi_bc} only
$a = R_{11}$ and $b=1/R_{22}^*$ are involved, the matrix elements 
of backscattering into the left electrode at $\mu_1=V+\mu$ and into
the right electrode at $\mu_2=\mu$. This reduction of the full
scattering matrix can be explained in a simple picture based
on the uncertainty relation: For long times $t \gg 1/V$, the
energy uncertainty $\Delta \epsilon$ of the scattered particles is smaller than
the voltage, $\Delta \epsilon \ll V$. The Fermi surface of the other electrode,
therefore, is not seen by the scattered particle, and only the
backscattering part of the scattering matrix is
of importance. For the left electrode, this is $R_{11}$, while
for the right electrode, the Pauli principle forbids virtual 
trajectories into the left electrode. After removal of these, the 
scattering amplitude becomes 
$R_{22} - R_{21} R_{11}^{-1} R_{12} = 1/R_{22}^*$.

Since $R_{11}$ and $R_{22}$ alone are not unitary 
the real phase shifts $\delta$ are replaced by complex
numbers. The implications, mainly, a broadening of the Fermi-edge
singularity, are shortly discussed below. For more details we refer 
to Refs.~\onlinecite{Muzykantskii03b,dAmbrumenil05}.

In the limit $|t|$ or $|t-t_f| \ll 1/V$, on the other hand, 
nonequilibrium constraints play no major role, and the equilibrium
expression \eqref{eq:RH_ND} is expected to become appropriate. 
This indeed is confirmed by the same argument of the 
uncertainty relation: For small times, $t \ll 1/V$, the energy
uncertainty $\Delta \epsilon$ of the scattered particles becomes large, 
$\Delta \epsilon \gg V$, and both Fermi surfaces at $\mu_1$ and $\mu_2$
are coupled. The full scattering matrix is involved.

\subsubsection{Low temperatures: $k_B T \ll V$}

Let us focus now on low temperatures, $k_B T \ll V$.
An approximate solution to $Y(t)$ to the RH problem can be
constructed in the same way as for the zero temperature
case: 
We decompose $R(t)$ according to Eq.~\eqref{eq:decomp_R},
and consider the RH problem for $\psi(t)$, described by
Eq.~\eqref{eq:psi_bc}. Since this problem is diagonal
with a time-independent boundary condition,
it can be exactly solved by the mapping
on the zero-temperature problem. Its solution is given by
Eq.~\eqref{eq:RH_YA} with $-\delta/\pi$ replaced by $\ln D/2 i\pi$,
\begin{equation}
	\psi(t) = 
	\exp\left(\frac{\ln D}{2\pi i}
	\ln\left[
		\frac{\sinh(\frac{\pi}{\beta}t)}{\sinh(\frac{\pi}{\beta}(t-t_f))}
	\right]\right).
\end{equation}
The function $W(t)$ is given by the same conditions 
\eqref{eq:def_W} as before, but where the regions
$s=1,\dots,4$ now are defined in the strip $\mathcal{S}_\beta$,
with additional branch cuts at $\mathrm{Im}\,t = \pm \beta/2$
(see Fig.~\ref{fig:zones} \mbox{(b)}) 
The validity of the approximation on the vertical branch cuts
remains unchanged from the zero temperature case, and corresponds
to $|t|,|t-t_f| \gg 1/V$.
On the additional horizontal cut, $W_2(t+i\beta/2)$ and $W_3(t-i\beta/2)$
differ in their nondiagonal parts, which are of the order of
$\e^{-\beta V/2}$. This leads to corrections for the identities
\eqref{eq:id+} and \eqref{eq:id-} of the form
\begin{equation}
\begin{aligned}
	f W_+ f = W_+ f + F_0 + \e^{-\beta V /2} F_+,\\
	f W_- f = f W_- + F_0 + \e^{-\beta V /2} F_-,
\end{aligned}
\end{equation}
where $F_0$ is the integral over all coinciding parts of $f W f$ 
along $\mathrm{Im}\,t = \pm \beta/2$. $F_+\neq F_-$ are the contributions
due to the mismatch of $W_2$ and $W_3$. Both, $F_+$ and $F_-$, are 
finite quantities of the order of unity. Since $\beta \gg 1/V$, their
contribution is negligible compared with the error arising from the 
vertical cuts.

On the other hand, for $|t-t_f|, |t| \ll 1/V \ll \beta$, the voltage
(i.e. the nonequilibrium situation) as well as the temperature become
unimportant. The solution to the RH problem, therefore, is that of
a zero temperature equilibrium system, expressed by Eq.~\eqref{eq:RH_ND}.

\subsubsection{High temperatures: $k_B T \gg V$}

At high temperatures, $k_B T \gg V$, we expect that $V$ plays 
no longer an important role. Indeed, since $\beta V \ll 1$,
the naive solution of the RH problem,
\begin{equation} \label{eq:naive_Y}
	Y(t) =
	\begin{pmatrix} 1 & \e^{i V t} \\ \e^{-i V t} & 1 \end{pmatrix}
	\left(
		\frac{\sinh(\frac{\pi}{\beta}t)}{\sinh(\frac{\pi}{\beta}(t-t_f))}
	\right)^{\delta/\pi}
\end{equation}
now becomes accurate. Because of the restriction of $t$ to the strip
$\mathcal{S}_\beta$, $Y(t)$ cannot diverge exponentially, so that $|Y(t)|$
remains bounded as $|t| \to \infty$ in $\mathcal{S}_\beta$. 
Yet, Eq.~\eqref{eq:naive_Y} violates the condition of periodicity,
$Y(t+i\beta/2) \neq Y(t-i\beta/2)$ due to the exponentials 
$\e^{\pm i V t}$. Again, this leads to different integrals $F$
from the integration along $t+i\beta/2$ and $t-i\beta/2$, respectively.
Since $\beta V \ll 1$, however, we can expand the exponential
functions, and we are led to similar identities as before
\begin{equation}
\begin{aligned}
	f Y_+ f = Y_+ f + \tilde{F}_0 + \beta V \tilde{F}_+,\\
	f Y_- f = f Y_- + \tilde{F}_0 + \beta V \tilde{F}_-,
\end{aligned}
\end{equation}
with $\tilde{F}_\pm$ of the order of unity.
The solution $Y$, and hence the inverse of the matrix, which is
solved by the RH problem, as well as the logarithm of the response
function are accurate within corrections of the
order of $\beta V$. 

\subsubsection{Nonequilibrium Fermi-edge singularities}

The results above allow us to extend the 
nonequilibrium Fermi-edge singularity problem to finite
temperatures. We consider the case where $R(t)$ has the 
form given in Eq.~\eqref{eq:R(t)} with $R=S=\e^{-2 i \delta}$.

The high temperature limit leads to an expression, which is 
unchanged from the equilibrium result \eqref{eq:sol_eq}.
At lower temperatures $k_B T \ll V$, on the other hand,
we can use the approximation $Y(t) \approx W(t)$ for 
for $|t|,|t-t_f|\gg 1/V$. With these functions and the 
decomposition of $R$ in Eq.~\eqref{eq:decomp_R}, 
the matrix products in Eq.~\eqref{eq:sol} can be evaluated
to
\begin{equation}
	\tr\left\{ \frac{\mathrm{d}Y_+}{\dt} Y_+^{-1}
	R^{-1} \frac{\mathrm{d}R}{\mathrm{d}\lambda} 
	\right\}
	=
	g(t) \frac{\mathrm{d}}{\dlambda}\frac{\tr\{(\ln D)^2\}}{2\pi i}
	- i V C(\lambda),
\end{equation}
where 
\begin{equation}
	g(t) = \frac{\mathrm{d}}{\dt} 
	\ln\left[
		\frac{\sinh(\frac{\pi}{\beta}(t+i0))}{\sinh(\frac{\pi}{\beta}(t-t_f+i0))}
	\right]
\end{equation}
and $C(\lambda)$ is a time-independent function of $\lambda$.
The term proportional to $C(\lambda)$ leads to a contribution to 
$\ln \chi^{(2)}_R$ that is linear in $V t_f$, and hence consists in a voltage 
dependent (generally complex) shift of the ground state energy, similar to 
that of $\ln\chi^{(1)}_R$. 
The time integration over $g(t)$ has to be cut off at $t=1/V$ and $t_f-t=1/V$,
where the solution $Y(t)$ crosses over into the equilibrium solution.
The remaining integral can be estimated by integrating the equilibrium
solution from $t=1/\xi$ to $1/V$.
This leads to 
\begin{multline}
	\ln\chi^{(2)}_R(t_f) 
	= -i t_f V C 
	- 
	\frac{\tr\{(\ln D)^2\}}{(2\pi)^2} 
	\ln\left[\frac{\sinh(\pi t_f/\beta)}{\sinh(\pi/\beta V)} \right]
	\\
	-
	\frac{\tr\{\delta^2\}}{\pi^2} 
	\ln\left[\frac{\sinh(\pi/\beta V)}{\sinh(\pi/\beta \xi)} \right],
\end{multline}
where we recall that $t_f \gg 1/V$.
Since $1/\xi \ll 1/V \ll \beta$, this can be rewritten as
\begin{multline}
	\ln\chi^{(2)}_R(t_f) 
	=
	-i t_f V C 
	-
	\frac{\tr\{(\ln D)^2\}}{(2\pi)^2} 
	\ln\left[\frac{\sinh(\pi t_f/\beta)}{\pi/\beta V} \right]
	\\
	-
	\frac{\tr\{\delta^2\}}{\pi^2} 
	\ln(\xi/V).
\end{multline}
For $1/\xi \ll t_f \ll 1/V$, this function crosses over into the 
equilibrium solution \eqref{eq:sol_eq}.

Since both $\delta$ and $\ln D$ are time-independent (for $t \in (0,t_f)$)
we see that, in this limit $k_B T\ll V$, the extension from zero to finite 
temperatures, originally achieved by the mapping $t \mapsto \tanh(\pi t/\beta)$, 
can here be obtained by a rule of thumb of replacing $t_f$ by 
$(\beta/\pi) \sinh(\pi t_f/\beta)$ in the logarithms. For $\chi_R(t)$ this 
means that the zero temperature power laws of the form $t_f^\alpha$ are 
replaced by $[\sinh(\pi t_f/\beta)]^\alpha$.

Therefore, finite temperatures replace the power-law tails by an exponential 
decay with a decay time proportional to the inverse temperature $\beta$.
In other words, while the system at zero temperature keeps a long-term memory 
to the initial perturbation, finite temperature fluctuations suppress it within 
a characteristic time set by $\beta/\alpha$.

The transition rate $p_{g\to e}(\Delta E)$ is determined by the Fourier 
integral \eqref{eq:trans_rate}.
An explicit calculation of the integral is possible and leads to a
Beta function, yet the following estimates are physically more instructive:

At zero temperature both, the singularity at $t\to 0$ as well as the power-law 
tails for $t \to \infty$ contribute equally to the integral. In the equilibrium situation,
we obtain the well-known Fermi-edge singularity, a power-law dependence on 
energy, $p_{g \to e} \sim \Delta E^{\tr\{\delta^2\}/\pi^2-1}$.
For $V \neq 0$, $\Delta E$ and the exponent, acquire an imaginary 
correction.
The nonequilibrium effects dominate for small energies $\Delta E < V$, and 
the imaginary corrections to $\Delta E$ lead to a broadening of the Fermi-edge 
singularity. 
The high energy tails, $\Delta E \gg V$, coincide with the equilibrium solution.

At finite but small temperatures, $k_B T \ll V$, the exponential decay of 
the response functions for $t \gg 1/k_B T$ causes an additional, Lorentzian, 
broadening of the singularity for $\Delta E < k_B T$. 
At high temperatures, $k_B T \gg V$, the nonequilibrium effects are completely 
washed out. For small energies, $\Delta E \ll k_B T$, the exponential decay 
then dominates the integral, and the singularity is broadened to a Lorentzian 
with a width proportional to $k_B T$. 
On the other hand, for high energies $\Delta E \gg k_B T$, the exponential 
decay is slow compared to the quickly oscillating phase factor. The behavior 
of the integral can be estimated through a saddle point approximation, leading 
to $p_{g \to e}(\Delta E) \sim \exp(-\Delta E/k_B T)$, corresponding to an
activated transition such as for a noninteracting system.

In summary, nonequilibrium constraints and temperature have the similar effect
of a broadening of the Fermi-edge singularity with two different shapes: 
A Lorentzian broadening from the temperature with a width given by $k_B T$, 
and a Lorentzian to the power of an expression depending on $\ln D$ from 
the nonequilibrium constraints with a width determined, to first order, by $V$. 
Away from the resonance, but still for $\Delta E \ll k_B T$, the transition 
rate has a power-law dependence on energy, which eventually turns into an 
activated behavior for $\Delta E > k_B T$.

% -----------------------------------------------------------------------------

\section{Conclusions}

We have shown that the RH approach developed in 
Refs.~\onlinecite{Muzykantskii03a} and \onlinecite{dAmbrumenil05} has an 
exact finite temperature extension. This finite temperature case can be
mapped onto a different zero temperature problem, for which much is
mathematically known\cite{singinteq}, and which allows, in some simpler
cases, an exact solution. Based on this equivalence we have 
constructed approximate solutions for the nonequilibrium RH problem
at finite temperatures.
These results provide a necessary tool for a general application 
as well as for further developments
of the RH approach and the study of Fermi-edge singularities
or related phenomena in nonequilibrium or nonadiabatic situations.

% -----------------------------------------------------------------------------

\begin{acknowledgments}
I thank D. E. Feldman, J. B. Marston, J. Merino, and T.-K. Ng for helpful
discussions and comments. This work is supported in part by the NSF under
grant number DMR-0213818.
\end{acknowledgments}

% -----------------------------------------------------------------------------
\appendix

% -----------------------------------------------------------------------------
\section{Time-representation of $R$}
\label{sec:time_repr_R}

The presented method requires that the time-representation of $R$, the matrix
elements of the many-body operator $\hat{R}$ between single-particle states,
is diagonal, $R = R(t)$.
For clarity and completeness, we here shortly state the conditions 
necessary for this, summarizing the arguments of Ref.~\onlinecite{dAmbrumenil05} 
(see also Refs.~\onlinecite{Muzykantskii03a} and, in particular, \onlinecite{Adamov01}). For a more extended
discussion we refer to the cited references.

In most cases of interest, $\hat{R}$ is connected to the evolution operators
$\hat{U}(t_f)$ and $\hat{U}_0(t_f)$ for the Hamiltonians $H$ and $H_0$, respectively.
For instance, in the case of the Fermi-edge singularities, $\hat{R}$ represents
the overlap between perturbed and unperturbed states, 
$\hat{R} = \hat{U}_0^\dagger(t_f) \hat{U}(t_f)$. 
On the other hand, for the shot noise spectrum of a tunneling barrier, $\hat{R}$ can be the 
moment generating function for the distribution of charge transfered out
of electrode $j=1$ during the time $t=0$ to $t_f$. 
If $\hat{Q}_1$ measures the charge of this electrode,
$\hat{R}(\lambda) = \hat{U}^\dagger(t_f) \e^{-i\lambda \hat{Q}_1} 
\hat{U}(t_f) \e^{i \lambda \hat{Q}_1}$.

Since the Hamiltonian is quadratic, the action of $\hat{U}(t_f)$ on the many-body
states can fully be described by its action on the single-particle states
(see also Appendix~\ref{sec:deriv_eq3}),
\begin{equation}
	\hat{U}(t_f) \mathbf{a}_{\epsilon'}^\dagger  |\rangle
	= \sum_{\epsilon} \e^{-i \epsilon t_f} \, \sigma(\epsilon,\epsilon') 
	\mathbf{a}_\epsilon^\dagger |\rangle,
\end{equation}
where $\sigma(\epsilon,\epsilon')$ is a $N \times N$ matrix in channel space, and
$|\rangle$ the true vacuum. 
The matrix $\sigma$ can be related to the single-particle scattering matrix $S(t,E)$
for a particle of energy $E$ calculated for the instantaneous value of the potential
$M(t)$. In general, this relation is complicated, but becomes simple when the following
condition of adiabaticity is met:
\begin{equation}
	\frac{\partial S^{-1}}{\partial t} \frac{\partial S}{\partial E} \ll 1.
\end{equation}
This equation encodes nothing than the condition that the scattering matrix
has to vary slowly during the characteristic scattering time of a particle 
(the Wigner delay time). In this case, the connection to the matrix $\sigma$ is
simple,
\begin{equation}
	\sigma_{j,j'}(\epsilon,\epsilon') = S_{j,j'}(\epsilon,\epsilon'),
\end{equation}
where
\begin{equation}
	S_{j,j'}(\epsilon,\epsilon') =
	\frac{1}{2\pi \sqrt{\nu_j \nu_{j'}}} \int \dt \, S_{j,j'}(t,E) \e^{i (\epsilon-\epsilon') t},
\end{equation}
with $E = (\epsilon+\epsilon')/2$ and $\nu_j$ the density of states in channel $j$.

The time representation $R(t)$ is obtained by a similar Fourier transform of 
$R_{j,j}(\epsilon,\epsilon') = \langle| a_{\epsilon,j} \hat{R} \, a_{\epsilon',j'}^\dagger |\rangle$.
\begin{equation}
	R_{j,j'}(\epsilon,\epsilon') =
	\frac{1}{2\pi \sqrt{\nu_j \nu_{j'}}} \int \dt \, R_{j,j'}(t,E) \e^{i (\epsilon-\epsilon') t}.
\end{equation}
As noted above, the definition of $R$ involves $\sigma$ or simple combinations
of $\sigma$, so that the same condition of adiabaticity applies.
The transformations in the RH method do not involve the dependence of these matrices on $E$.
At low temperatures and in equilibrium, the relevant physics is restricted to energies
close to the chemical potential $\mu$, so that we can fix $E = \mu$. For nonequilibrium
situations, the treatment remains valid as long as the dependence of $R$ on $E$ is weak.

% -----------------------------------------------------------------------------
\section{Derivation of Eq.~(\ref{eq:det_chi})}
\label{sec:deriv_eq3}

In this appendix, we demonstrate the equality 
\begin{equation} \label{eq:many_single}
	\Tr[ \hat{\rho} \hat{R} ] 
	= \det\bigl( f R + (1-f) \bigr),
\end{equation}
stated in Eq.~\eqref{eq:det_chi}
that connects a many-body description on the left-hand side
to a single-body representation on the right-hand side.
$\hat{R}$ is the many-body operator of interest, $\hat{\rho}$
the initial many-body density matrix at $t=0$, 
$\hat{\rho} \propto \e^{-\beta \hat{H}_0}$,
and $\Tr$ the trace over all many-body states.
On the other hand, $R = \{ R_{ij}\}$ is the $N_s \times N_s$ matrix between
the $N_s$ possible single-particle states in the system, 
$R_{ij} = \langle| a_i \hat{R}  a_j^\dagger|\rangle$ (for simplicity of notation, 
we here label single-particle states by a single index $i$ only -- in the main
text this is split into energy $\epsilon$ and channel indices $j$).
$|\rangle$ is the true vacuum, not the ground state, $f$ the corresponding Fermi 
function, and $\det$ the determinant over the $N_s \times N_s$ matrices.

The Hamiltonian $\hat{H}$ is quadratic in the single-particle operators for all times $t$.
At time $t=0$ and at zero temperature, the density matrix projects on the
ground state $|0\rangle$, which is a Slater determinant in the single-particle states 
$|\psi_i\rangle = a_i^\dagger |\rangle$. If the many-body operator 
$\hat{R}$ conserves particle numbers, it is, therefore, fully described 
by its action on the single-particle constituents of the ground state,
represented by the matrix $R$.
Explicitly, for a system of $N \le N_s$ particles, the ground state is 
$|0\rangle = (N!)^{-1/2} \sum_{\pi} (-1)^\pi |\psi_{\pi(1)}\rangle \otimes \dots
\otimes |\psi_{\pi(N)}\rangle$, where the sum runs over all permutations $\pi$,
$(-1)^\pi$ is the sign of the permutation, and where we assume that $i=1,\dots,N$
labels the $N$ single-particle states of lowest energy.
We have 
$\hat{R} |\psi_i\rangle = \hat{R}_{i} |\psi_i\rangle = \sum_{j=1}^{N_s} R_{ji} |\psi_j\rangle$,
where $\hat{R}_i$ is the restriction of $\hat{R}$ to the single-particle state $i$.
Further 
$\hat{R} (|\psi_1\rangle \otimes |\psi_2\rangle \otimes \dots ) 
= \hat{R}_1 |\psi_1\rangle \otimes \hat{R}_2 |\psi_2\rangle \otimes \dots$, 
from which follows that
\begin{equation}
\begin{aligned}
	\langle 0 | \hat{R} | 0 \rangle
	&= \frac{1}{N!} 
	\sum_{\pi \pi'} (-1)^{\pi+\pi'}
	\Bigl[ \langle\psi_{\pi'(1)}| \otimes \dots \otimes \langle\psi_{\pi'(N)}| \Bigr]
  \\&\ 
	\times
	\hat{R}
	\Bigl[|\psi_{\pi(1)}\rangle \otimes \dots \otimes |\psi_{\pi(N)}\rangle \Bigr]
\\
	&= \frac{1}{N!} 
	\sum_{\pi \pi'} (-1)^{\pi+\pi'}
	\Bigl[ \langle\psi_{\pi'(1)}| \otimes \dots \otimes \langle\psi_{\pi'(N)}| \Bigr]
  \\&\ 
	\times
	\Bigl[
	\sum_{j_1,\dots,j_N=1}^{N_s} R_{j_1\pi(1)} |\psi_{j_1}\rangle \otimes \dots 
	\otimes R_{j_N \pi(N)}|\psi_{j_N}\rangle \Bigr]
\\
	&= \frac{1}{N!} 
	\sum_{\pi \pi'} (-1)^{\pi+\pi'}
	R_{\pi'(1)\pi(1)} 
	\dots
	R_{\pi'(N)\pi(N)} 
\\
	&= \sum_{\pi} (-1)^{\pi}
	R_{1\pi(1)} \dots R_{N\pi(N)}.
\end{aligned}
\end{equation}
This is the determinant of $R$, taken over the occupied states only, and thus
corresponds to the right-hand side of Eq.~\eqref{eq:many_single}.

The generalization to nonzero temperatures is immediate: 
Every excited state of $N$ particles yields a similar determinant of $R$ for 
a different selection of $N$ particles $\psi_j$. The density matrix $\hat{\rho}$ 
weights these states with the Boltzmann factors $\propto \e^{-\beta \hat{H}_0}$.
Since these are exponentials of single-particle operators, they just provide
the weights $\e^{-\beta \epsilon_j}$ for the occupied single-particle states
that sum up to the Fermi function when summing over all excited states.
Therefore, Eq.~\eqref{eq:many_single} remains valid at nonzero temperatures.

% -----------------------------------------------------------------------------
\vfill

% -----------------------------------------------------------------------------

% -----------------------------------------------------------------------------

% -----------------------------------------------------------------------------

\end{document}